\documentclass[aps,a4paper,superscriptaddress,preprintnumbers,showpacs,amsmath,amssymb]{revtex4}
\usepackage{graphicx}
\usepackage{dcolumn}
\usepackage{color}
\usepackage{latexsym,amsfonts}
\usepackage{textcomp}
\usepackage{bm}

\usepackage{ulem}

\begin{document}

\title{Non--Relativistic Approximation \\of the Dirac Equation for Slow
Fermions in Static Metric Spacetimes}

\author{A. N. Ivanov}\email{ivanov@kph.tuwien.ac.at}
\author{M. Pitschmann}\email{pitschmann@kph.tuwien.ac.at}
\affiliation{Atominstitut, Technische Universit\"at Wien, Stadionallee
  2, A-1020 Wien, Austria}

\date{\today}

\begin{abstract}
We analyse the non--relativistic approximation of the Dirac equation
for slow fermions moving in spacetimes with a static metric, caused by
the weak gravitational field of the Earth and a  chameleon
field, and derive the most general effective gravitational potential,
induced by a static metric of spacetime.  The derivation of the
non--relativistic Hamilton operator of the Dirac equation is carried
out by using a standard Foldy--Wouthuysen (SFW) transformation. We
discuss the chameleon field as source of a torsion field and
torsion--matter interactions.
\end{abstract}
\pacs{03.65.Pm, 04.25.-g, 04.25.Nx, 14.80.Va}

\maketitle

\section{Introduction}
\label{sec:introduction}

The non--relativistic quantum states of (ultra)cold neutrons in the
gravitational field of the Earth above a totally reflecting mirror
were found in
\cite{1975AmJPh..43...25G,wallis1992trapping,1999AmJPh..67..776G} and
measured in
\cite{nesvizhevsky2002quantum,Nesvizhevsky:2003ww,nesvizhevsky2005study,westphal2007quantum},
while the transitions between quantum gravitational states of
ultracold neutrons were measured in \cite{abele2010ramsey}. A possible
explanation for the acceleration of our Universe at the present time
is given by assuming the existence of a so-called chameleon field
\cite{khoury2004chameleon,mota2007evading,PhysRevLett.97.151102,waterhouse2006introduction}.
The presence of such a field would modify the gravitational potential
$U_g(\vec{r}\,) = \vec{g}\cdot \vec{r}$ above a mirror
\cite{brax2011strongly} and between two mirrors
\cite{ivanov2013influence} as
\begin{eqnarray}\label{eq:1}
  U(\vec{r}\,) = \vec{g}\cdot \vec{r} + \frac{\beta}{M_{\rm
      Pl}}\,\phi(\vec{r}\,)\,,
\end{eqnarray}
where $\phi(\vec{r}\,)$ is a chameleon field, $\beta$ is the
chameleon--matter coupling constant
\cite{khoury2004chameleon,mota2007evading,PhysRevLett.97.151102,waterhouse2006introduction,brax2011strongly}
and $M_{\rm Pl} = 1/\sqrt{8\pi G_N} = 2.435\times 10^{27}\,{\rm eV}$
is a Planck mass, expressed in terms of the gravitational constant
$G_N$ \cite{beringer2012review}. The gravitational potential energy of
(ultra)cold neutrons is related to the potential $U(\vec{r}\,)$ via
$\Phi(\vec{r}\,) = m\,U(\vec{r}\,)$, where $m$ is the neutron mass.

For the analysis of the Dirac equation in the gravitational field we
employ the static metric $ds^2 = g_{\mu\nu}(x) dx^{\mu} dx^{\nu} =
g_{00}(x) dt^2 + g_{ij}(x)dx^idx^j = V^2 dt^2 - W^2 (d\vec{r}\,)^2$,
where $x^{\mu} = (t, \vec{r}\,)$. The components $g_{00}(x) = V^2$ and
$g_{ij}(x) = - W^2\, \delta_{ij}$ of the metric tensor $g_{\mu\nu}(x)$
are functionals of the potential $U(\vec{r}\,)$. The dynamics of
(ultra)cold neutrons moving in such a spacetime is defined by the
Dirac equation
\begin{eqnarray}\label{eq:2}
  i\,\frac{\partial \psi(t,\vec{r}\,)}{\partial t} = \hat{H}(\vec{r},\vec{\nabla},\vec{\sigma}\,)\,\psi(t,\vec{r}\,)\,,
\end{eqnarray}
where $\vec{\nabla}$ is the gradient with respect to $\vec{r}$ and
$\vec{\sigma}$ are the $2\times 2$ Pauli matrices \cite{Itzykson:1980rh}. Using
tetrad fields \cite{PhysRevD.23.2157}, the Hamilton operator
$\hat{H}(\vec{r},\vec{\nabla},\vec{\sigma}\,)$ of the Dirac equation
in coordinate representation takes the form
\cite{obukhov2001spin}
\begin{eqnarray}\label{eq:3}
  \hat{H}(\vec{r},\vec{\nabla},\vec{\sigma}\,) = \gamma^0mV - i\,\frac{V}{W}\,\gamma^0\vec{\gamma}\cdot\Big(\vec{\nabla} + \frac{\vec{(\nabla}V)}{2 V}
+ \frac{\vec{(\nabla}W)}{W}\Big)\,,
\end{eqnarray}
where $\gamma^{\mu} = (\gamma^0, \vec{\gamma}\,)$ are the Dirac
matrices in Minkowski spacetime \cite{Itzykson:1980rh}. 
The matrix elements of the Hamilton operator are given by
\begin{eqnarray}\label{eq:4}
  \langle f|\hat{H}|i\rangle = \int d^3x\,\sqrt{-\det(g_{ij})}\,\psi_f^\dagger(t, \vec{r}\,)\hat{H}\,\psi_i(t, \vec{r}\,)\,,
\end{eqnarray}
where $\sqrt{-\det(g_{ij})} = W^3$. In order to deal with a Hermitian Hamilton operator it is convenient to redefine $\psi(t, \vec{r}\,) = \{-\det(g_{ij})\}^{-1/4}\Psi(t, \vec{r}\,) = W^{-3/2}\Psi(t, \vec{r}\,)$ to obtain
\begin{eqnarray}\label{eq:5}
  \langle f|\hat{H}|i\rangle = \int d^3x\,\Psi_f^\dagger(t, \vec{r}\,)\hat{\bf H}\,\Psi_i(t, \vec{r}\,)\,,
\end{eqnarray}
which defines the Hermitian Hamilton operator $\hat{\bf H} = W^{3/2}\hat{H}W^{-3/2}$ \cite{PhysRevD.23.2157}
\begin{eqnarray}\label{eq:6}
  \hat{\bf H} = \gamma^0mV - i\,\frac{V}{W}\,\gamma^0\vec{\gamma}\cdot\vec{\nabla} - \frac{i}{2}\,\gamma^0\vec{\gamma}\cdot\Big(\vec{\nabla}\frac{V}{W}\Big)\,.
\end{eqnarray}
This operator agrees with the one given by Obukhov
\cite{obukhov2001spin}.

The paper is organised as follows. In section \ref{sec:sfw} we derive
the Pauli equation as the non--relativistic limit of the Dirac
equation describing the interaction of (ultra)cold neutrons with the
gravitational field of the Earth and a chameleon field. In section
\ref{sec:conclusion} we discuss the obtained results.

\section{Standard Foldy--Wouthuysen transformation}
\label{sec:sfw}

In order to obtain a non-relativistic Hamilton operator we follow
Foldy and Wouthuysen \cite{foldy1950dirac} and decompose the Hamilton
operator Eq.~(\ref{eq:6}) into ``even'' and ``odd'' operators and
delete the ``odd'' operators by employing unitary transformations.

The first term in Eq.~(\ref{eq:6}), proportional to $\gamma^0$, is an
``even'' operator, whereas the second and third terms, proportional to
$\gamma^0 \vec{\gamma}$, are ``odd'' operators. For the elimination of
the ``odd'' operators we perform the unitary transformation
\begin{eqnarray}\label{eq:7}
  \hat{\bf H}_1 = e^{\,i \hat{S}_1}\hat{\bf H}e^{\,- i\hat{S}_1} =
  \hat{\bf H} + i[\hat{S}_1,\hat{\bf H}] +
  \frac{i^2}{2!}\,[\hat{S}_1,[\hat{S}_1,\hat{\bf H}]] + \ldots\,,
\end{eqnarray}
with the operator $\hat{S}_1$ given by
\begin{eqnarray}\label{eq:8}
  \hat{S}_1 = - \frac{i}{2m}\,\gamma^0 \Big(-\frac{1}{W}\,i\gamma^0
  \vec{\gamma}\cdot \vec{\nabla} -
  \frac{i}{2V}\,\gamma^0\vec{\gamma}\cdot\Big(\vec{\nabla}\frac{V}{W}\Big)\Big)
  = -\frac{1}{2mW}\,\vec{\gamma}\cdot\vec{\nabla} -
  \frac{1}{4mV}\,\vec{\gamma}\cdot\Big(\vec{\nabla}\frac{V}{W}\Big)\,.
\end{eqnarray}
For the expansion of the Hamilton operator $\hat{\bf H}_1$ in
Eq.~(\ref{eq:7}) we neglect terms of order $\mathcal O(1/m^2)$.  The
expressions of the commutators $[\hat{S}_1,\hat{\bf H}]$ and
$[\hat{S}_1,[\hat{S}_1,\hat{\bf H}]]$ are given by
\begin{eqnarray}\label{eq:9}
  i[\hat{S}_1,\hat{\bf H}] &=&
  \frac{i}{2W}\,\gamma^0\vec{\gamma}\cdot(\vec{\nabla}V) +
  i\,\frac{V}{W}\,\gamma^0\vec{\gamma}\cdot \vec{\nabla} -
  \frac{1}{2m}\,\gamma^0\,\Big(\vec{\nabla}\frac{V}{W^2}\Big)\cdot\vec{\nabla}
  + \frac{i}{2m}\,\gamma^0\,\Big(\vec{\nabla}\frac{V}{W^2}\Big)
  \cdot(\vec{\Sigma}\times \vec{\nabla}\,) -
  \frac{V}{mW^2}\,\gamma^0\Delta
  \nonumber\\ &&-\,\frac{1}{2mW}\,\gamma^0\,\Big(\Delta\,\frac{V}{W}\Big)
  -
  \frac{1}{mW}\,\gamma^0\,\Big(\vec{\nabla}\,\frac{V}{W}\Big)\cdot\vec{\nabla}
  +
  \frac{i}{2}\,\gamma^0\vec{\gamma}\cdot\Big(\vec{\nabla}\,\frac{V}{W}\Big)
  +
  \frac{1}{4mVW}\,\gamma^0(\vec{\nabla}\,V)\cdot\Big(\vec{\nabla}\,\frac{V}{W}\Big)
  \nonumber\\ &&-\,\frac{i}{4mW^3}\,\gamma^0\,\vec{\Sigma}\cdot\Big((\vec{\nabla}\,V)
  \times (\vec{\nabla}\,W)\Big) -
  \frac{1}{4mV}\,\gamma^0\,\Big(\vec{\nabla}\,\frac{V}{W}\Big)^2\,,
\end{eqnarray}
as well as
\begin{eqnarray}\label{eq:10}
  \frac{i^2}{2!}\,[\hat{S}_1,[\hat{S}_1,\hat{\bf H}]] &=&
  -\frac{1}{8mW^3}\,\gamma^0(\vec{\nabla}W)\cdot(\vec{\nabla}V) -
  \frac{i}{4mW^3}\,\gamma^0\,\vec{\Sigma}\cdot\Big((\vec{\nabla}\,W)
  \times (\vec{\nabla}\,V)\Big) +
  \frac{1}{4mW^2}\,\gamma^0\,(\vec{\nabla}V)\cdot\vec{\nabla}
  \nonumber\\ &&+\,\frac{1}{4m}\,\gamma^0\,\Big(\vec{\nabla}\frac{V}{W^2}\Big)
  \cdot\vec{\nabla} -
  \frac{i}{4m}\,\gamma^0\,\Big(\vec{\nabla}\frac{V}{W^2}\Big)\cdot(\vec{\Sigma}
  \times \vec{\nabla}) + \frac{V}{2mW^2}\,\gamma^0\Delta +
  \frac{1}{4mW}\,\gamma^0\,\Big(\Delta\,\frac{V}{W}\Big)
  \nonumber\\ &&+\,\frac{1}{2mW}\,\gamma^0\,\Big(\vec{\nabla}\,\frac{V}{W}\Big)
  \cdot\vec{\nabla} +
  \frac{1}{8mV}\,\gamma^0\,\Big(\vec{\nabla}\,\frac{V}{W}\Big)^2 +
  \frac{1}{8mW^2}\,\gamma^0(\Delta V)\,.
\end{eqnarray}
As result, we obtain the Hamilton operator $\hat{\bf H}_1$ 
\begin{eqnarray}\label{eq:11}
  \hat{\bf H}_1 &=& \gamma^0mV +
  \frac{i}{2W}\,\gamma^0\vec{\gamma}\cdot(\vec{\nabla}V) -
  \frac{1}{4m}\,\gamma^0\,\Big(\vec{\nabla}\frac{V}{W^2}\Big)\cdot\vec{\nabla}
  +
  \frac{i}{4m}\,\gamma^0\,\Big(\vec{\nabla}\frac{V}{W^2}\Big)\cdot(\vec{\Sigma}\times
  \vec{\nabla}\,) - \frac{V}{2mW^2}\,\gamma^0\Delta
  \nonumber\\ &&-\,\frac{1}{4mW}\,\gamma^0\,\Big(\Delta\,\frac{V}{W}\Big)
  -
  \frac{1}{2mW}\,\gamma^0\,\Big(\vec{\nabla}\,\frac{V}{W}\Big)\cdot\vec{\nabla}
  +
  \frac{1}{4mVW}\,\gamma^0(\vec{\nabla}\,V)\cdot\Big(\vec{\nabla}\,\frac{V}{W}\Big)
  - \frac{1}{8mV}\,\gamma^0\,\Big(\vec{\nabla}\,\frac{V}{W}\Big)^2
  \nonumber\\ &&-\,\frac{1}{8mW^3}\,\gamma^0(\vec{\nabla}W)\cdot(\vec{\nabla}V)
  + \frac{1}{4mW^2}\,\gamma^0\,(\vec{\nabla}V)\cdot\vec{\nabla} +
  \frac{1}{ 8mW^2}\,\gamma^0(\Delta V)\,.
\end{eqnarray}
After the first unitary transformation, the resulting Hamilton
operator $\hat{\bf H}_1$, calculated to order $1/m$, still contains an
``odd'' operator proportional to $\gamma^0\vec{\gamma}$. This ``odd''
operator we remove by employing a second unitary transformation
\begin{eqnarray}\label{eq:12}
  \hat{\bf H}_2 = e^{\,i \hat{S}_2}\hat{\bf H}_1e^{\,- i\hat{S}_2} = 
  \hat{\bf H}_1 + i[\hat{S}_2,\hat{\bf H}_1] + \frac{i^2}{2!}\,[\hat{S}_2,[\hat{S}_2,\hat{\bf H}_1]] + \ldots\,,
\end{eqnarray}
with operator $\hat{S}_2$ given by
\begin{eqnarray}\label{eq:13}
  \hat{S}_2 = - \frac{i}{2m}\,\gamma^0 \Big( \frac{i}{2VW}\,\gamma^0 \vec{\gamma}\cdot \vec{\nabla}V\Big) =
  \frac{1}{4mVW}\,\vec{\gamma}\cdot \vec{\nabla}V\,.
\end{eqnarray}
Neglecting terms of order $\mathcal O(1/m^2)$ we obtain
\begin{eqnarray}\label{eq:14}
  i[\hat{S}_2,\hat{\bf H}_1] &=& - \frac{i}{2W}\,\gamma^0\vec{\gamma}\cdot(\vec{\nabla}V) - \frac{1}{4mVW^2}\,\gamma^0(\vec{\nabla}V)^2\,, \nonumber\\
  \frac{i^2}{2!}\,[\hat{S}_2,[\hat{S}_2,\hat{\bf H}_1]] &=& \frac{1}{8mVW^2}\,\gamma^0(\vec{\nabla}V)^2\,.
\end{eqnarray}
As result, the Hamilton operator $\hat{\bf H}_2$ is given by 
\begin{eqnarray}\label{eq:15}
  \hat{\bf H}_2 &=& \gamma^0mV - \frac{1}{4m}\,\gamma^0\,\Big(\vec{\nabla}\frac{V}{W^2}\Big)\cdot\vec{\nabla} + \frac{i}{4m}\,\gamma^0\,\Big(\vec{\nabla}\frac{V}{W^2}\Big)\cdot(\vec{\Sigma}\times \vec{\nabla}\,) - \frac{V}{2mW^2}\,\gamma^0\Delta - \frac{1}{4mW}\,\gamma^0\,\Big(\Delta\,\frac{V}{W}\Big) \nonumber\\
  && -\,\frac{1}{2mW}\,\gamma^0\,\Big(\vec{\nabla}\,\frac{V}{W}\Big)\cdot\vec{\nabla} + \frac{1}{4mVW}\,\gamma^0(\vec{\nabla}\,V)\cdot\Big(\vec{\nabla}\,\frac{V}{W}\Big) - \frac{1}{8mV}\,\gamma^0\,\Big(\vec{\nabla}\,\frac{V}{W}\Big)^2 - \frac{1}{8mW^3}\,\gamma^0(\vec{\nabla}W)\cdot(\vec{\nabla}V)  \nonumber\\
  &&+\,\frac{1}{4mW^2}\,\gamma^0\,(\vec{\nabla}V)\cdot\vec{\nabla} + \frac{1}{8mW^2}\,\gamma^0(\Delta V) - \frac{1}{8mVW^2}\,\gamma^0(\vec{\nabla}V)^2\,,
\end{eqnarray}
containing only ``even'' operators. After these two unitary transformations, the Dirac equation reads
\begin{eqnarray}\label{eq:16}
  i\,\gamma^0 \frac{\partial\Psi(t,\vec{r}\,)}{\partial t} &=& \bigg\{mV - \frac{1}{4m}\,\Big(\vec{\nabla}\frac{V}{W^2}\Big)\cdot\vec{\nabla} + \frac{i}{4m}\,\Big(\vec{\nabla}\frac{V}{W^2}\Big)\cdot(\vec{\Sigma}\times \vec{\nabla}\,) - \frac{V}{2mW^2}\,\Delta - \frac{1}{4mW}\,\Big(\Delta\,\frac{V}{W}\Big) \nonumber\\
  && -\,\frac{1}{2mW}\,\Big(\vec{\nabla}\,\frac{V}{W}\Big)\cdot\vec{\nabla} + \frac{1}{4mVW}\,(\vec{\nabla}\,V)\cdot\Big(\vec{\nabla}\,\frac{V}{W}\Big) - \frac{1}{8mV}\,\Big(\vec{\nabla}\,\frac{V}{W}\Big)^2 - \frac{1}{8mW^3}\,(\vec{\nabla}W)\cdot(\vec{\nabla}V)  \nonumber\\
  &&+\,\frac{1}{4mW^2}\,(\vec{\nabla}V)\cdot\vec{\nabla} + \frac{1}{8mW^2}\,(\Delta V) - \frac{1}{8mVW^2}\,(\vec{\nabla}V)^2\bigg\}\,\Psi(t,\vec{r}\,)\,,
\end{eqnarray}
with the four-component spinor $\Psi(t,\vec{r}\,)$. As was shown by Foldy and Wouthuysen \cite{foldy1950dirac}, in the low-energy approximation of the Dirac equation the ``large'' and ``small'' components of the Dirac bispinor wavefunctions can be separated by the projection operators $(1 + \gamma^0)/2$ and $(1 - \gamma^0)/2$, respectively. Next, we introduce the two-component spinor $\varphi(t,\vec{r}\,)$ as
\begin{eqnarray}\label{eq:17}
\left(\begin{array}{c}
  \varphi(t,\vec{r}\,) \\
  0 \\
  \end{array}\right)
 = \frac{1 + \gamma^0}{2}\,e^{\,im t}\,\Psi(t,\vec{r}\,)\,,
\end{eqnarray}  
where the phase $e^{\,imt}$ removes the highly oscillating mode of the two-component spinor. Multiplying Eq.~(\ref{eq:16}) with the projection operator $(1 + \gamma^0)/2$, we obtain the Pauli equation 
\begin{eqnarray}\label{eq:18}
  i\,\frac{\partial\varphi(t,\vec{r}\,)}{\partial t} = \Big(-\frac{1}{2m}\,\Delta + \Phi(\vec{r},\vec{\nabla},\vec{\sigma}\,)\Big)\,\varphi(t,\vec{r}\,)\,,
\end{eqnarray}
with the potential $\Phi(\vec{r}, \vec{\nabla}, \vec{\sigma}\,)$ given by 
\begin{eqnarray}\label{eq:19}
  \Phi(\vec{r}, \vec{\nabla}, \vec{\sigma}\,) &=& m(V - 1) - \frac{1}{4m}\,\Big(\vec{\nabla}\frac{V}{W^2}\Big)\cdot\vec{\nabla} + \frac{i}{4m}\,\Big(\vec{\nabla}\frac{V}{W^2}\Big)\cdot(\vec{\sigma}\times \vec{\nabla}\,) + \frac{W^2 - V}{2mW^2}\,\Delta - \frac{1}{4mW}\,\Big(\Delta\,\frac{V}{W}\Big) \nonumber\\
  && -\,\frac{1}{2mW}\,\Big(\vec{\nabla}\,\frac{V}{W}\Big)\cdot\vec{\nabla} + \frac{1}{4mVW}\,(\vec{\nabla}\,V)\cdot\Big(\vec{\nabla}\,\frac{V}{W}\Big) - \frac{1}{8mV}\,\Big(\vec{\nabla}\,\frac{V}{W}\Big)^2 - \frac{1}{8mW^3}\,(\vec{\nabla}W)\cdot(\vec{\nabla}V)  \nonumber\\
  &&+\,\frac{1}{4mW^2}\,(\vec{\nabla}V)\cdot\vec{\nabla} + \frac{1}{8mW^2}\,(\Delta V) - \frac{1}{8mVW^2}\,(\vec{\nabla}V)^2\,.
\end{eqnarray}
This is the most general effective low-energy gravitational potential
of slow fermions, induced by a static spacetime metric $ds^2 = V^2
dt^2 - W^2 (d\vec{r}\,)^2$, which has ever been calculated previously
in the literature. The effective gravitational potential
Eq.~(\ref{eq:19}) can be used to any order of perturbation theory with
respect to the potential $U(\vec{r}\,)$.

For the derivation of the effective gravitational potential in a weak
gravitational and chameleon field and comparison to other results,
obtained in the literature, we use $V^2 = 1 + 2 U(\vec{r}\,)$ and $W^2
= 1 - 2 \gamma\,U(\vec{r}\,)$ and expand the potential
Eq.~(\ref{eq:19}) to linear order in $U(\vec{r}\,)$. This gives
\begin{eqnarray}\label{eq:20}
  \Phi(\vec{r}, \vec{\nabla}, \vec{\sigma}\,) = m\,U(\vec{r}\,) -
  \frac{1 + 2\gamma}{2m}\,\Big(U(\vec{r}\,)\,\Delta + (\vec{\nabla}U(\vec{r}\,))\cdot
  \vec{\nabla} +  \frac{1}{4}\,(\Delta
  U(\vec{r}\,)) -
  \frac{i}{2}\,(\vec{\nabla}U(\vec{r}\,))\cdot(\vec{\sigma}\times
  \vec{\nabla}\,) \Big)\,.
\end{eqnarray}
Using the Newtonian gravitational potential $U(\vec{r}\,) = -G_NM/r$, where $G_N$ and $M$ are the gravitational coupling constant and mass, respectively, we arrive at the effective gravitational potential 
\begin{eqnarray}\label{eq:21}
  \Phi(\vec{r}, \vec{\nabla}, \vec{\sigma}\,) = -G_N\frac{Mm}{r} + (1
  + 2\gamma)\,\frac{G_NM}{2m}\,\Big( \frac{1}{r}\,\Delta -
  \frac{\vec{r}}{r^3}\cdot \vec{\nabla} - \pi\,\delta^{(3)}(\vec{r}\,)
  + \frac{1}{2} \frac{\vec{\sigma}\cdot\hat{\vec{L}}}{r^3} \Big)\,,
\end{eqnarray}
where $\hat{\vec{L}} = \vec{r}\times\hat{\vec{p}}$ is the orbital
momentum operator and $\Delta(1/r) = - 4\pi\,
\delta^{(3)}(\vec{r}\,)$. This result agrees with the effective
gravitational potential obtained by Fischbach, Freeman and Cheng
\cite{PhysRevD.23.2157} and recently by Jentschura and Noble
\cite{jentschura2013nonrelativistic,jentschura2014foldy}.

Using the potential $U(\vec{r}\,)$ given by Eq.~(\ref{eq:1}), we
obtain the potentials $\Phi_{\rm
  G}(\vec{r},\vec{\nabla},\vec{\sigma}\,)$ and $\Phi_{\rm
  Ch}(\vec{r},\vec{\nabla},\vec{\sigma}\,)$, caused by gravity and a
chameleon field, respectively:
\begin{eqnarray}\label{eq:22}
  \Phi_{\rm G}(\vec{r}, \vec{\nabla}, \vec{\sigma}\,) &=&
  \vec{g}\cdot\bigg(m\,\vec{r} - \frac{1 + 2 \gamma}{2m}\,\Big[
    \vec{r}\,\Delta + \vec{\nabla} - \frac{i}{2}\,(\vec{\sigma}\times
    \vec{\nabla}\,)\Big]\bigg)\,, \nonumber\\ \Phi_{\rm Ch}(\vec{r},
  \vec{\nabla}, \vec{\sigma}\,) &=& \frac{\beta}{M_{\rm
      Pl}}\bigg(m\,\phi(\vec{r}\,) - \frac{1 +
    2\gamma}{2m}\,\Big[\phi(\vec{r}\,)\,\Delta +
    (\vec{\nabla}\phi(\vec{r}\,))\cdot\vec{\nabla} +
    \frac{1}{4}\,(\Delta\phi(\vec{r}\,)) -
    \frac{i}{2}\,(\vec{\nabla}\phi(\vec{r}\,))\cdot(\vec{\sigma}
    \times\vec{\nabla}\,) \Big]\bigg)\,.
\end{eqnarray}
These effective gravitational and chameleon field potentials should be
employed for the experimental analysis of the fine structure of
quantum gravitational states of ultracold neutrons in qBounce
experiments and properties of a chameleon field
\cite{abele2010ramsey,abele2009qubounce,Jenke2009318,abele2012gravitation,Jenke:2014yel}.

For the subsequent analysis it is convenient to represent the obtained
results in the following form
\begin{eqnarray}\label{eq:23}
  {\rm \hat{H}} = {\rm \hat{H}}_0 + \hat{V}_{\rm G} + \hat{V}_{\rm Ch}
  + \hat{V}_{\cal T}\,,
\end{eqnarray}
where the operators ${\rm \hat{H}}_0$, $\hat{V}_{\rm G}$, $\hat{V}_{\rm Ch}$ and $\hat{V}_{\cal T}$ are defined by
\begin{eqnarray}\label{eq:24}
 {\rm \hat{H}}_0 &=& - \frac{1}{2m}\,\Delta + m \vec{g}\cdot
 \vec{r}\,,\nonumber\\ \hat{V}_{\rm G} &=& - \frac{1 + 2
   \gamma}{2m}\,\vec{g}\cdot (\vec{r}\,\Delta +
 \vec{\nabla}\,)\,,\nonumber\\ \hat{V}_{\rm Ch}
 &=&\beta\,\frac{m}{M_{\rm Pl}}\,\phi(\vec{r}\,) - \beta\,\frac{1 +
   2\gamma}{2 m M_{\rm Pl}}\,\Big(\phi(\vec{r}\,)\,\Delta +
 (\vec{\nabla}\phi(\vec{r}\,))\cdot\vec{\nabla} +
 \frac{1}{4}\,(\Delta\phi(\vec{r}\,))\Big)\,,\nonumber\\ \hat{V}_{\cal
   T} &=& i\frac{1 + 2\gamma}{4 m}\,\Big(\vec{g} + \frac{\beta}{M_{\rm
     Pl}}\,\vec{\nabla}\phi(\vec{r}\,)\Big)\cdot (\vec{\sigma}\times
 \vec{\nabla}\,)\,.
\end{eqnarray}
In this case the operators $\hat{V}_{\rm G}$, $\hat{V}_{\rm Ch}$ and
$\hat{V}_{\cal T}$ are determined on the class of wavefunctions
$\varphi_n(\vec{r}\,)$, which are eigenfunctions of the operator ${\rm
  H}_0$, i.e.
\begin{eqnarray}\label{eq:25}
{\rm \hat{H}}_0 \varphi_n(\vec{r}\,) = E_n \varphi_n(\vec{r}\,)\,,
\end{eqnarray}
where $E_n$ are the binding energies of ultracold neutrons and $n =
1,2, \ldots$ is the principle quantum number.

\section{Torsion--matter like interaction}
\label{sec:torsion1}

In this section we show that the effective potential $\hat{V}_{\cal
  T}$ describes an effective torsion--neutron interaction. This can be
inferred from the comparison to the results obtained by Kostelecky
{\it et al.}  \cite{Kostelecky:2007kx}. For this aim we rewrite
$\hat{V}_{\cal T}$ in the relativistic covariant form
\begin{eqnarray}\label{eq:26}
\delta {\cal L}_{\cal T}(x) =
\frac{i}{2}\,g_{\cal T}\,{\cal T}_{\mu}(x)\bar{\psi}(x)\,\sigma^{\mu\nu}
\overleftrightarrow{\partial_{\nu}}\psi(x)\,,
\end{eqnarray}
where $\psi(x)$ is the neutron field operator and
$A(x)\overleftrightarrow{\partial_{\nu}}B(x) = A(x) \partial_{\nu}B(x)
- (\partial_{\nu}A(x)) B(x) $ and $\sigma^{\mu\nu} =
\frac{i}{2}(\gamma^{\mu}\gamma^{\nu} - \gamma^{\nu} \gamma^{\mu})$ is
one of the Dirac matrices \cite{Itzykson:1980rh}. From comparison of 
Eq.~(\ref{eq:26}) to Eq.~(2) in Ref.~
\cite{Kostelecky:2007kx} we obtain
\begin{eqnarray}\label{eq:27}
g_{\cal T}{\cal T}_{\mu}(x) &=& \xi^{(5)}_6\,T_{\mu}(x) +
\xi^{(5)}_7\,A_{\mu}(x)\,,
\end{eqnarray}
where $\xi^{(5)}_6$ and $\xi^{(5)}_7$ are phenomenological coupling
constants, introduced by Kostelecky {\it et
  al.} \cite{Kostelecky:2007kx} for the description of a
torsion--Dirac--fermion field interaction. The fields $T_{\mu}(x)$ and
$A_{\mu}(x)$ are related to the torsion tensor field
${T^{\alpha}}_{\mu\nu}(x) = - {T^{\alpha}}_{\nu\mu}(x)$ as $T_{\mu}(x)
= g^{\alpha\beta}T_{\alpha\beta\mu}(x)$ and $A_{\mu}(x) =
\frac{1}{6}\varepsilon^{\alpha\beta\gamma\mu}T_{\alpha\beta\gamma}(x)$,
respectively, where $g^{\alpha\beta}$ and
$\varepsilon^{\alpha\beta\gamma\mu}$ are the inverse metric tensor and
Levi--Civita tensor in Minkowski spacetime. The part of the effective
Lagrangian Eq.~(\ref{eq:26}), which can be expressed in terms of the
potential $\hat{V}_{\cal T}$ in the non--relativistic limit, is
\begin{eqnarray}\label{eq:28}
\delta {\cal L}_{\cal T}(x) = i\,g_{\cal T}\vec{\cal T}\cdot
\bar{\psi}(x)(\vec{\Sigma}\times \vec{\nabla}\,)\psi(x) + \ldots =
i\,g_{\cal T} \vec{\cal T}\cdot \varphi^{\dagger}(x)(\vec{\sigma}\times
\vec{\nabla}\,)\varphi(x) + \ldots\,,
\end{eqnarray}
where $\varphi(x)$ is the operator of the large component of the Dirac
bispinor field operator $\psi(x)$.  From Eq.~(\ref{eq:28}) and
Eq.~(\ref{eq:24}) for $\hat{V}_{\cal T}$ we obtain
\begin{eqnarray}\label{eq:29}
g_{\cal T}\vec{\cal T}(x) = - \frac{1 + 2\gamma}{4 m}\,\Big(\vec{g} +
\frac{\beta}{M_{\rm Pl}}\,\vec{\nabla}\phi(\vec{r}\,)\Big)\,.
\end{eqnarray}
According to Colladay and Kostelecky, the constant part of a torsion
field should be removed by the redefinition of the wavefunction of a
fermion field given in Eq.~(30) in
Ref.~\cite{Colladay:1998fq}. Contrary to that, the part of a torsion
field depending on spacetime coordinates cannot be removed in general
\cite{Kostelecky:2003fs}. In our approach a spacetime dependent part
is given by the contribution of a chameleon field. 

In the following we analyse a redefinition of the wavefunction of the
matter--field of ultracold neutrons moving in a spatial region
restricted in the $z$--direction by either a mirror from below or
confined by two mirrors while unrestricted in the $(x,y)$ plane.  In
our approach, a redefinition of the wavefunctions of ultracold neutrons
corresponds to a rearrangement of the terms in the Hamilton operator
${\rm \hat{H}}$, given in Eq.~(\ref{eq:24}):
\begin{eqnarray}\label{eq:30}
{\rm \hat{H}} = {\rm \hat{H}}'_0 + \hat{V}_{\rm G} + \hat{V}_{\rm Ch}
+ \hat{V}'_{\cal T}\,,
\end{eqnarray}
where we have denoted
\begin{eqnarray}\label{eq:31}
{\rm \hat{H}}'_0 &=& - \frac{1}{2m}\,\Delta + m \vec{g}\cdot \vec{r} +
i\frac{1 + 2\gamma}{4 m}\,\vec{g}\cdot (\vec{\sigma}\times
\vec{\nabla}\,)\,,\nonumber\\ \hat{V}'_{\cal T} &=& i\frac{1 +
  2\gamma}{4 m} \frac{\beta}{M_{\rm
    Pl}}\,\vec{\nabla}\phi(\vec{r}\,)\cdot (\vec{\sigma}\times
\vec{\nabla}\,)\,.
\end{eqnarray}
Thus, we propose to determine the operators $\hat{V}_{\rm G}$,
$\hat{V}_{\rm Ch}$ and $\hat{V}'_{\cal T}$ on the class of
wavefunctions $\varphi'(\vec{r}\,)$, which are eigenfunctions of the
Hamilton operator ${\rm \hat{H}}'_0$
\begin{eqnarray}\label{eq:32}
i\frac{\partial\varphi'(\vec{r},t)}{\partial t} = {\rm \hat{H}}'_0 \varphi'(\vec{r},t)\,.
\end{eqnarray}
The removal of the constant part of the torsion field can be carried out by employing a unitary transformation. 
For this aim we analyse the Pauli equation Eq.~(\ref{eq:32}) in the following form
\begin{eqnarray}\label{eq:33}
i\frac{\partial \varphi'(\vec{r},t)}{\partial t} =\Big( -
\frac{1}{2m}\,\Delta + m \vec{g}\cdot \vec{r} + i\frac{1 + 2\gamma}{4
  m}\, (\vec{g}\times \vec{\sigma}\,)\cdot
\vec{\nabla}\,\Big)\varphi'(\vec{r},t)\,,
\end{eqnarray}
where we have used the Hamilton operator ${\rm \hat{H}}'_0$, given in Eq.~(\ref{eq:31}). Performing a unitary transformation
\begin{eqnarray}\label{eq:34}
\varphi'(\vec{r},t) = \Omega\,\tilde{\varphi}(\vec{r},t)\,,
\end{eqnarray}
we transcribe the Pauli equation Eq.~(\ref{eq:33}) into the form
\begin{eqnarray}\label{eq:35}
i\frac{\partial \tilde{\varphi}(\vec{r},t)}{\partial t} &=& \Big( -
\frac{1}{2m}\,\Delta + m \vec{g}\cdot \vec{r} - \frac{1}{2m}\,\Omega^{-1}(\Delta
\Omega) + i\frac{1 + 2\gamma}{4 m}\,
\Omega^{-1}(\vec{g}\times \vec{\sigma}\,)\cdot
(\vec{\nabla}\Omega)\Big)
\tilde{\varphi}(\vec{r},t) \nonumber\\
 &+&\Big(- \frac{1}{m}\,\Omega^{-1}(\vec{\nabla}\Omega) + i\frac{1 + 2\gamma}{4 m}\,
\Omega^{-1}(\vec{g}\times \vec{\sigma}\,)\,\Omega\Big) \cdot
\vec{\nabla}\tilde{\varphi}(\vec{r},t)\,.
\end{eqnarray}
For the removal of the required constant part of the torsion field we set
\begin{eqnarray}\label{eq:36}
\vec{\nabla} \Omega = i\frac{1 + 2\gamma}{4}\,(\vec{g}\times
\vec{\sigma}\,)\,\Omega\,,
\end{eqnarray}
which reduces Eq.~(\ref{eq:35}) to  
\begin{eqnarray}\label{eq:37}
i\frac{\partial \tilde{\varphi}(\vec{r},t)}{\partial t} &=& \Big( -
\frac{1}{2m}\,\Delta + m \vec{g}\cdot \vec{r} - \frac{1}{2m}\,\Omega^{-1}(\Delta
\Omega) + i\frac{1 + 2\gamma}{4 m}\,
\Omega^{-1}(\vec{g}\times \vec{\sigma}\,)\cdot
(\vec{\nabla}\Omega)\Big)
\tilde{\varphi}(\vec{r},t)\,.
\end{eqnarray}
Using Eq.~(\ref{eq:36}) we transcribe the last two terms on the right--hand side (r.h.s.) of Eq.~(\ref{eq:37}) as
\begin{eqnarray}\label{eq:38}
 - \frac{1}{2m}\,\Omega^{-1}(\Delta
\Omega) + i\frac{1 + 2\gamma}{4 m}\,
\Omega^{-1}(\vec{g}\times \vec{\sigma}\,)\cdot
(\vec{\nabla}\Omega) = -
g^2\, \frac{(1 + 2\gamma)^2}{16 m}\,.
\end{eqnarray}
Hence, the wavefunction $\tilde{\varphi}(\vec{r},t)$ satisfies
the Schr\"odinger equation
\begin{eqnarray}\label{eq:39}
i\frac{\partial \tilde{\varphi}(\vec{r},t)}{\partial t} = \bigg( -
\frac{1}{2m}\,\Delta + m \vec{g}\cdot \vec{r} - g^2\, \frac{(1 +
  2\gamma)^2}{16 m}\bigg)\,\tilde{\varphi}(\vec{r},t)\,.
\end{eqnarray}
For stationary quantum states $\tilde{\varphi}(\vec{r},t) = \tilde{\varphi}(\vec{r}\,)\,e^{\,-iE_nt}$ we obtain
\begin{eqnarray}\label{eq:40}
\tilde E_n\tilde{\varphi}(\vec{r}\,) = \Big( -
\frac{1}{2m}\,\Delta + m \vec{g}\cdot \vec{r}\Big)\,\tilde{\varphi}(\vec{r}\,)\,.
\end{eqnarray}
where $\tilde E_n = E_n + g^2\,(1 + 2\gamma)^2/16 m$. Such a constant shift of energy levels  
does not appear in the transition frequencies $\omega_{nk} = \tilde E_n - \tilde E_k = E_n - E_k$, which are 
experimentally observed \cite{abele2010ramsey,abele2009qubounce,Jenke2009318,abele2012gravitation,Jenke:2014yel}.
Hence, we have shown that the constant part of the torsion can be
removed from the interaction to any order in perturbation theory. To
linear order the solution of Eq.~(\ref{eq:36}) is
equal to
\begin{eqnarray}\label{eq:41}
 \Omega = 1 + i\frac{1 + 2\gamma}{4}\,(\vec{g}\times
 \vec{\sigma}\,)\cdot \vec{r}\,.
\end{eqnarray}
This agrees with the transformations analysed by Kostelecky
\cite{Kostelecky:2007kx,Colladay:1998fq,Kostelecky:2003fs}.

\section{Analysis of observability of constant part of torsion field }
\label{sec:torsion2}

As has been pointed out by Kostelecky
\cite{Kostelecky:2007kx,Colladay:1998fq,Kostelecky:2003fs} the
constant parts of the torsion fields, removed by a redefinition of the
wavefunctions of fermions, should not be observable to first order
perturbation theory. In this section we analyse the problem of
observability of the constant part of the torsion field $g_{\cal
  T}\vec{{\cal T}}$, which we removed from the potential operator
$\hat{V}_{\rm T}$ by a redefinition of the wavefunctions of ultracold
neutrons $\varphi'_n(\vec{r}\,) =
\Omega\,\tilde{\varphi}_n(\vec{r}\,)$. For this aim we calculate the
contributions of the effective gravitational potential $\hat{V}_{\rm
  G}$ to the energy levels of quantum gravitational states of
ultracold neutrons. To first in
order perturbation theory the observables are diagonal matrix elements
of the potential operator $\hat{V}_{\rm G}$, i.e. $\langle
n|\hat{V}_{\rm G}|n\rangle$ given by (see \cite{Jenke:2014yel})
\begin{eqnarray}\label{eq:42}
\langle n|\hat{V}_{\rm G}|n\rangle = \frac{\displaystyle \int d^3x \,
  \varphi'^{\dagger}_n(\vec{r}\,)\hat{V}_{\rm G}\varphi'_n(\vec{r}\,)
}{\displaystyle \int d^3x\, \varphi'^{\dagger}_n(\vec{r}\,)
  \varphi'_n(\vec{r}\,) } = \frac{\displaystyle \int d^3x \,
  \tilde{\varphi}^{\dagger}_n(\vec{r}\,)\Omega^{-1}\hat{V}_{\rm
    G}\Omega \tilde{\varphi}_n(\vec{r}\,) }{\displaystyle \int d^3x\,
  \tilde{\varphi}^{\dagger}_n(\vec{r}\,) \tilde{\varphi}_n(\vec{r}\,)
}\,.
\end{eqnarray}
Setting $\varphi'_n(\vec{r}\,) = \Omega\,\tilde{\varphi}_n(z)$ and
$\vec{g} = g\,\vec{e}_z$, using Eq.~(\ref{eq:40}) and skipping
standard intermediate calculations we arrive at the result
\begin{eqnarray}\label{eq:43}
\langle n|\hat{V}_{\rm G}|n\rangle = (1 + 2 \gamma)\,E_ng \int
dz\,z\,|\tilde{\varphi}_n(z)|^2 - (1 + 2 \gamma)\,mg^2 \int
dz\,z^2\,|\tilde{\varphi}_n(z)|^2 + (1 + 2\gamma)^3\frac{g^3}{8m}\int
dz\,z\,|\tilde{\varphi}_n(z)|^2 \,,
\end{eqnarray}
where the integration over $z$ should be carried out in accord to the
experimental setup of the qBounce experiments
\cite{abele2010ramsey,abele2009qubounce,Jenke2009318,abele2012gravitation,Jenke:2014yel}. 

The obtained result shows that the contributions of the constant part
of the torsion field to the energy levels and, correspondingly, to the
transition frequencies \cite{Jenke:2014yel} are of third order in $g$. 
This confirms Kostelecky's
assertion \cite{Kostelecky:2007kx,Colladay:1998fq,Kostelecky:2003fs}
that the constant part of the torsion field is unobservable to
first order in perturbation theory.  

\section{Conclusion}
\label{sec:conclusion}

We have analysed the non--relativistic approximation of the Dirac
equation for (ultra)cold neutrons, propagating in a spacetime with
static metric $ds^2 = g_{\mu\nu}(x)dx^{\mu}dx^{\nu} = V^2dt^2 -
W^2(d\vec{r}\,)^2$.  The components of the static metric,
$g_{00}(\vec{r}\,) = V^2$ and $g_{ij}(\vec{r}\,) = -
W^2\,\delta_{ij}$, are functionals of the potential $U(\vec{r}\,) =
\vec{g}\cdot \vec{r} + (\beta/M_{\rm Pl})\,\phi(\vec{r}\,)$, where
$\vec{g}\cdot \vec{r}$ and $(\beta/M_{\rm Pl})\,\phi(\vec{r}\,)$ are
the contributions of the gravitational field of the Earth and a
chameleon field, respectively.

We have carried out the non--relativistic reduction of the Dirac
equation rigorously by employing a standard Foldy--Wouthuysen (SFW)
transformation neglecting terms of order $\mathcal{O}(1/m^2)$.  We
have found the effective potential as a functional of the metric
components $V$ and $W$ in Eq.~(\ref{eq:19}).  For $V^2 = 1 + 2
U(\vec{r}\,)$ and $W^2 = 1 - 2 \gamma\,U(\vec{r}\,)$, we have derived
the effective gravitational potential in Eq.~(\ref{eq:20}) to linear
order in $U(\vec{r}\,)$. In addition to the gravitational field, our
expression contains the contribution of a chameleon field.

After the redefinition of the wavefunctions of fermions we arrive at the effective potential
\begin{eqnarray}\label{eq:44}
  \hat{V}_{\rm eff} = \hat{V}_{\rm G} + \vec{V}_{\rm Ch} + \hat{V}'_{\cal T}\,,
\end{eqnarray}
where the last term corresponds to a torsion--matter interaction
\cite{Kostelecky:2007kx,Colladay:1998fq,Kostelecky:2003fs,hehl1976general,shapiro2002physical,hammond2002torsion}
with the torsion field $g_{\cal T} \vec{\cal T} = - ((1 + 2
\gamma)/4m)(\beta/M_{\rm Pl})\vec{\nabla}\phi(\vec{r}\,)$.  The
obtained result shows that in principle a chameleon field serves as a
source for a torsion field interaction with matter fields.  The
possibility for a torsion field interaction to be induced by a scalar
field was discussed by Hojman {\it et al.}  \cite{hojman1978gauge}
(see also \cite{hammond2002torsion}). Following Kostelecky
\cite{Kostelecky:2007kx,Colladay:1998fq,Kostelecky:2003fs} we have
removed the constant part of the torsion field by a redefinition of
the wavefunctions of fermions.  Furthermore, we have shown that such a
change of the wavefunctions of ultracold neutrons in the qBounce
experiments does not lead to any observable contributions to first
order in perturbation theory. A more detailed investigation of the
subtleties involved in the derivation of the non-relativistic limit of
the Dirac equation for slow fermions in a gravitational field as well
as in a chameleon field as a possible source for a torsion field
interaction we are planning to perform in our forthcoming publication.

Finally, we would like to note that the first two terms of the
effective potential Eq.~(\ref{eq:20}), namely $m\, U(\vec{r}\,) -
((1+2\gamma)/2m)\, U(\vec{r}\,)\, \Delta$, reproduce the effective
non--relativistic Hamilton operator $ H^{(0,1)}_{\rm NR}$ of
spin--independent low--energy interactions of Dirac fermions moving in
a curved spacetime with an asymptotically flat and diagonal metric
tensor (see Eq.~(62) of Ref.~\cite{Kostelecky:2010ze}). The absence of
derivative terms of the metric tensor in the Hamilton operator $
H^{(0,1)}_{\rm NR}$ is justified in \cite{Kostelecky:2010ze} by the
smallness of spacetime variations of the metric tensor with respect to
its deviation from the Minkowski spacetime metric tensor.

\section{Acknowledgement}

We are grateful to Hartmut Abele and Philippe Brax for their interest
in our work and discussions, to Ulrich Jentschura for discussions of
properties of Foldy--Wouthuysen transformations and to Alan Kostelecky
for fruitful discussions concerning properties of a torsion field and
its interaction with matter.

This work was supported by the Austrian
``Fonds zur F\"orderung der Wissenschaftlichen Forschung'' (FWF) under
the contracts I862-N20 (AI) and I689-N16 (MP).

\bibliographystyle{h-physrev3.bst}
\bibliography{ChamDP}
\end{document}